\documentclass[fleqn,usenatbib]{mnras}

\usepackage{newtxtext,newtxmath}
\usepackage[T1]{fontenc}
\DeclareRobustCommand{\VAN}[3]{#2}
\let\VANthebibliography\thebibliography
\def\thebibliography{\DeclareRobustCommand{\VAN}[3]{##3}\VANthebibliography}

\usepackage{graphicx}
\usepackage{amsmath}
\usepackage{amssymb}
\usepackage{xcolor}
\usepackage{changes}
\definechangesauthor[name={RB}, color=purple]{RB}

\title[Prospects of detecting a connection between stellar-mass BBH mergers and AGN]
{Detectability of a spatial correlation between stellar-mass black hole mergers
and Active Galactic Nuclei in the Local Universe}

\author[N. Veronesi et al.]{
Niccolò Veronesi,$^{1}$\thanks{E-mail: veronesi@strw.leidenuniv.nl}
Elena Maria Rossi,$^{1}$
Sjoert van Velzen,$^{1}$
Riccardo Buscicchio$^{2,3}$
\\
$^{1}$Leiden Observatory, Leiden University, PO Box 9513, 2300 RA Leiden, The Netherlands\\
$^{2}$Dipartimento di Fisica ``G. Occhialini'', Universit\'a degli Studi di Milano-Bicocca,
Piazza della Scienza 3, 20126 Milano, Italy\\
$^{3}$INFN, Sezione di Milano-Bicocca, Piazza della Scienza 3, 20126 Milano, Italy
}

\date{Accepted 2022 May 6. Received 2022 May 4; in original form 2022 March 14}

\pubyear{2022}

\begin{document}
\label{firstpage}
\pagerange{\pageref{firstpage}--\pageref{lastpage}}
\maketitle

\begin{abstract}
\noindent The origin of the Binary Black Hole (BBH) mergers detected through Gravitational Waves (GWs) by the
LIGO-Virgo-KAGRA (LVK) collaboration remains debated. One fundamental reason is our ignorance of their host
environment, as the typical size of an event's localization volume can easily contain thousands of galaxies.
A strategy around this is to exploit statistical approaches to assess the spatial correlation between these
mergers and astrophysically motivated host galaxy types, such as Active Galactic Nuclei (AGN).
We use a Likelihood ratio method to infer the degree of GW-AGN connection out to $z=0.2$. 
We simulate BBH mergers whose components' masses are sampled from a realistic distribution of the underlying
population of Black Holes (BHs). Localization volumes for these events are calculated assuming two different
interferometric network configurations.
These correspond to the configuration of the third (O3) and of the upcoming fourth (O4) LVK observing runs.
We conclude that the 13 BBH mergers detected during the third observing run at $z\leq0.2$ are not enough to reject with
a \(3\sigma\) significance the hypothesis according to which there is no connection between GW and AGN more luminous than
$\approx 10^{44.3}\rm{erg}\ \rm{s}^{-1}$, that have number density higher than \(10^{-4.75}\textrm{Mpc}^{-3}\).
However, 13 detections are enough to reject this no-connection hypothesis when rarer categories of AGN are considered,
with bolometric luminosities greater than $\approx 10^{45.5}\rm{erg}\ \rm{s}^{-1}$.
We estimate that O4 results will potentially allow us to test fractional contributions to the total BBH
merger population from AGN of any luminosity higher than \(80\%\).
\end{abstract}



\section{Introduction}

Since the first direct GW detection has been announced 
\citep{abbott16}, the two interferometers of Advanced LIGO \citep{ligo2015} and 
the one of Advanced Virgo \citep{acernese15} have measured the signal coming from
tens of compact objects mergers in three observing runs
\citep{abbott19,abbott21,abbott21duepuntouno,abbott21gwtc3cat}. Thanks to improved
sensitivities and the addition of a fourth detector, KAGRA \citep{somiya12,aso13},
this number will grow in the upcoming years \citep{abbott18}.

Different formation pathways for these merging BBHs have been
proposed \citep{mapelli21}. They might arise from the evolution of
isolated stellar binary systems \citep{dominik12,belczynski16,spera19}, or form
in dense environments, in which dynamical interactions can efficiently drive
binaries of compact objects towards the merger
\citep{stone17a,rodriguez18,antonini19,gerosa19,rodriguez21,rizzuto21}. One particular example
of such environments can be the accretion disk around Super Massive Black Holes (SMBHs) in
Active Galactic Nuclei \citep{bartos17b,stone17b,mckernan18,ford19,samsing20,gayathri21}. It has been shown
that in such an environment, compact objects can migrate towards a radius close to the
one of the innermost stable circular orbit, and there be trapped for the remaining AGN lifetime 
\citep{peng21}. The large number density of compact objects and the high escape velocity in that
region can facilitate the occurrence of  hierarchical mergers (i.e. mergers in which at least one of
the two components is the remnant of a previous merger) \citep{yang19,gerosa21,wang21hierar}.
The mass of the remnants of hierarchical mergers can be higher than the lower bound of the Pair Instability
mass gap predicted by stellar formation models \citep{farmer19,woosley21}.
This formation pathway has therefore the theoretical advantage of being able to explain the
non-vanishing merger rate inferred for binaries with components heavier than \(50 \rm{M}_{\odot}\)
\citep{ligo21pop}.

There are potentially several ways to address the formation pathways' open question and in
particular to assess a plausible connection between GW events (BBH mergers in particular) and AGN.
The most straightforward would be to directly detect an ElectroMagnetic (EM) counterpart in coincidence
with the GW event. This might be possible in dense environments like the accretions disks of AGN
\citep{mckernan19,wang21a}, and such a counterpart might have already been observed \citep{graham20}
(\citealp[However, see also][]{ashton21}). 
The typical localization volumes of GW events make their association with an EM counterpart challenging.
The interferometers currently operating are in fact only able to associate to GW detections comoving volumes
that can easily contain thousands of different galaxies. 
Similarly to what happens in the case of the emission of an EM counterpart, a companion GW signal
can be originated from the same source of a detected event in the case of mergers happening near a SMBH.
These events could therefore be identified by the independent detection of an associated gravitational echo
\citep{kocsis13,gondan21}.

Another way to infer the origin of the detected events is by statistically comparing the
measured source population properties, such as mass and spin distributions, with model expectations.
This kind of analysis has been done for several potential host environments, including AGN
\citep{mckernan20,gayathri21,tagawa21,wang21hierar,guopeng22}. While the expected distribution of spin parameters is
still debated, all these analyses conclude that heavy (\(\geq50M_{\odot}\)) stellar-mass BHs are
expected to be generated through the AGN formation channel.

Finally, the increasing number of detections allows us to exploit statistical approaches to explore
the spatial correlation between GW events and specific types of host environments. These approaches
can overcome the big challenge of large localization volumes.
\citet{bartos17} proposed a statistical likelihood-ratio-based method to find out how many GW
detections would be needed to establish which fraction of BBH mergers detected through GWs happened in an
AGN. This earlier work was based on the GW localization volume distribution expected for detections performed
by the LIGO-Virgo network at design sensitivity and assuming only mergers of pairs of \(10\rm{M}_\odot\) BHs.\\
In this work, we present an analysis based on the same method, although we use simulated GW detections constructed
from the latest results on the inference of the underlying BBH component masses' distribution. To simulate these
detections we employ detectors' sensitivities representative of the third observing run of the LIGO and Virgo
interferometers, as well as those expected to characterize the fourth one, when KAGRA will join the network.

This paper is organized as follows: In Section \ref{sec:method} we provide an overview of all the steps
of the analysis, with details in the following subsections. How we simulated the GW detections
the localization volumes of which are needed in the statistical analysis is described in subsection
\ref{sec:simul}, while in subsection \ref{sec:estimation} we present how this statistical 
investigation works. The results of our works are presented ins section \ref{sec:res}.
Finally, in Section \ref{sec:concl} we draw conclusions and discuss the next steps to improve our it to
observed data.

\vspace{-0.5em}
\section{Method}
\label{sec:method}

To investigate the spatial correlation between Gravitational Waves 90\%
Credibility Level localization volumes (thereafter "V90")
and the positions of AGN in the local Universe, we first build two catalogues of simulated
GW detections anchored in current observations.
For the first, we simulate the response of the detector network active during O3.
For the second catalogue, we use the same synthetic population of BBHs, and we simulate their
detection by the interferometric network configuration expected for O4, which includes also KAGRA.
To create the simulated detections we first sample the joint probability
distribution of the binary mass ratio \(q=m_2/m_1\) and primary component's mass \(m_1\); which is,
by definition, greater than the mass of the secondary one, \(m_2\).
We then sample the spin distribution for each binary component, the distribution for the inclination
of the orbital plane with respect to the line of sight, and for the luminosity distance between
the position of the event and the detectors. The assumed distributions, as well as the configurations
and the detector sensitivity curves used in our simulations, are described in section \ref{sec:simul}.

Once the mock observations have been simulated, we evaluate V90 for all the
detections using BAYESTAR \citep{singer16}, a sky localization algorithm able to perform in a
few seconds a Bayesian, non-Markov Chain Monte Carlo analysis.

We then use the newly created distribution for V90 to sample a set of comoving volumes that are
then exploited in an algorithm based on the likelihood-ratio method described in \citet{bartos17}.
This algorithm crossmatches the positions of the GW localization volumes with the positions
of AGN in the local Universe, which are assumed to be isotropically distributed in comoving volume.
The final output of this algorithm (described in detail in \ref{sec:estimation}) is the number of
GW detections needed to test the hypothesis according to which a certain fraction
(\(f_{\rm{agn}}\)) of the detected BBH mergers happened in an AGN; having the chance of rejecting the
no-correlation hypothesis (none of the detected BBH mergers happened in an AGN) with a
given confidence.

\subsection{Simulation of GW detections}
\label{sec:simul}

A distribution of V90 is required by our statistical method. We obtain such distribution by simulating
several realistic GW detections for both O3 and O4 configurations. We describe the details of these
simulations in the following sections.

\subsubsection{Source population}
\label{sec:underlying}

Our simulated GW events are derived from the population analysis based on the latest results of the LVK
Collaboration.
We assume for \(m_1\) the {\scshape Power Law + Peak} analytical model presented in \citet{abbott21pop}
and we simultaneously sample values of \(m_1\) and \(q\) from their joint posterior probability distribution.
This distribution has been obtained through the standard hierarchical bayesian analysis presented in 
\citet{ligo21pop} and posterior samples are publicly available. The secondary component's mass is then
calculated as \(m_2=qm_1\).
We use the same mass distribution to simulate BBH mergers irrespective of them happening in an AGN.
This is done to maintain our estimate conservative and model-independent. We, therefore, neglect the
effects of the hypotheses according to which GW events originated in dense environments are more likely
to involve higher-mass BHs with respect to the ones that originated from an isolated binary system.

For simplicity, we assume for all the BHs the spin direction to be aligned with respect to the binary orbital
angular momentum, and a uniform spin magnitude distribution between \(0\) and \(1\). The distribution of V90 is not
expected to be significantly affected by such an assumption.

The simulated binaries are uniformly distributed in comoving volume and their inclinations \(\iota\) are
sampled according to a uniform distribution over \(\arccos(\iota)\).
The cosmological parameters we assume during our analysis are the ones inferred from the Planck Cosmic
Microwave Background observations \citep{planck18}.

\subsubsection{The network of detectors}
Next, we simulate the network of detectors. The whole analysis presented in this work is done for
two different settings: the first one aims at reproducing the V90 distribution for O3, while the
second one aims to forecast the distribution of the detected volumes expected during O4. 
In both cases, we assume a duty cycle of 78\% for all the different detectors individually,
and we keep a network Signal-to-Noise Ratio (SNR) threshold of \(8\), adding a Gaussian
measurement error to the SNR and requiring that at least two detectors contribute to the network SNR with an
individual SNR \(\geq4\).
The signals of the injected events are then compared with the detectors' noise in the $10-5000$Hz
frequency range. We use an IMRPhenomD waveform type \citep{husa16,khan16} to model the injections.

To reproduce the volume distribution of the events measured during O3, we model a network
of three detectors: LIGO Hanford, LIGO Livingstone, and Virgo, using the sensitivities characterized
by the following IDs: {\scshape aLIGOMidLowSensitivityP1200087} for the two LIGO interferometers, and
{\scshape AdVMidLowSensitivityP1200087} for Virgo interferometer.

For the O4 predictions, we add a fourth KAGRA-like interferometer, and we change the
sensitivity curves of each detector. Specifically, we use {\scshape aLIGOAdVO4T1800545} for LIGO and Virgo
detectors, and {\scshape aLIGOKAGRA80MpcT1800545} for KAGRA.

\subsubsection{Evaluation of 90\% CL localization volumes}

\begin{figure}
    \centering
    \includegraphics[width=0.49\textwidth]{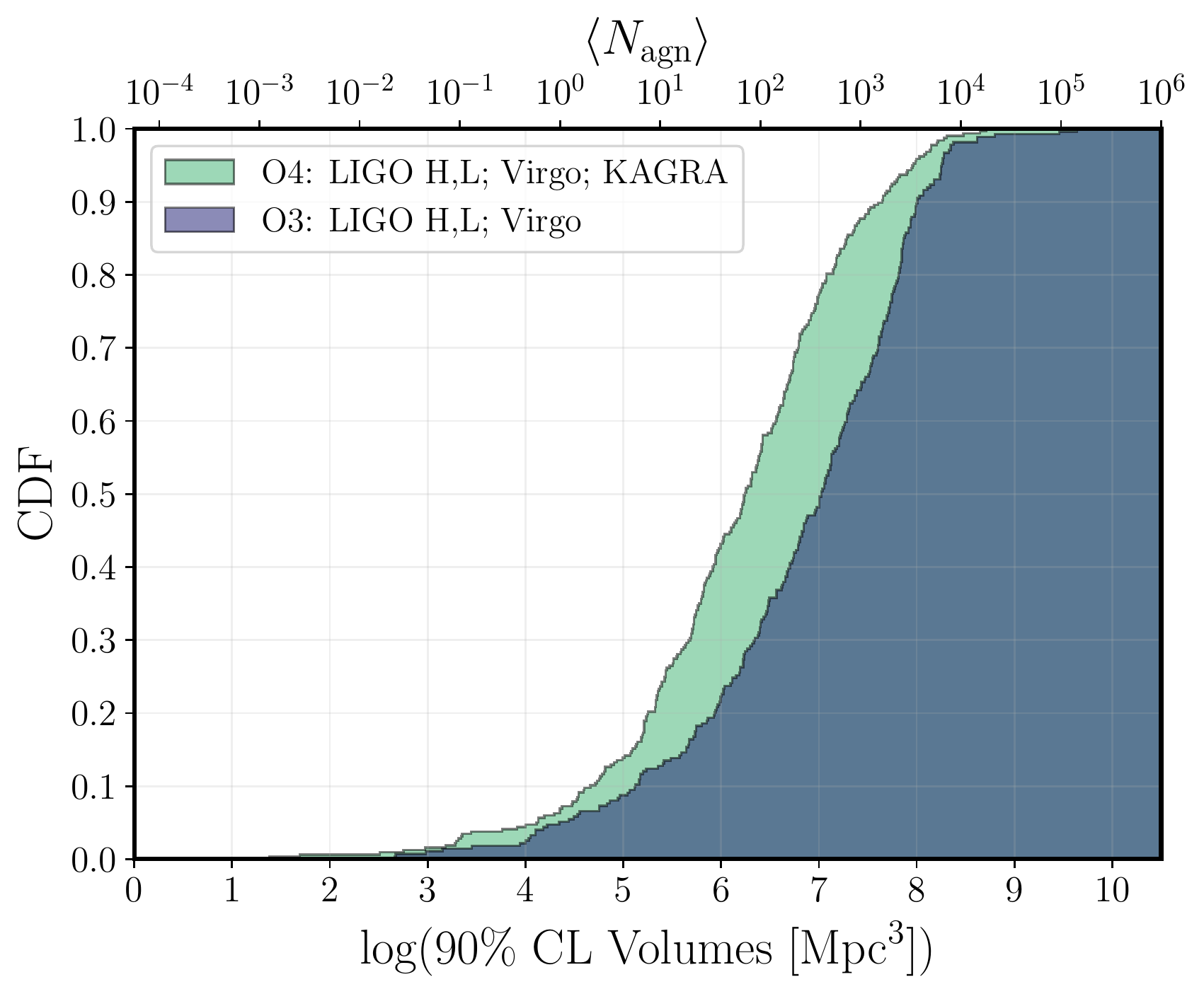}
    \caption{Cumulative distributions of the 90\% CL localization volumes of simulated GW events with SNR\(>8\) and
    \(z\leq0.2\). The blue and the green histograms are for O3 and O4 runs, respectively. The top axis shows the expected
    number of AGN within the corresponding localization volume, for a homogeneous distribution of AGN with a number
    density of \(n_{\rm{agn}}=10^{-4.75} \textrm{Mpc}^{-3}\).}
    \label{fig:cdfvol}
\end{figure}

For O3 (O4), out of the 200k (100k) injections, 663 (1737) have a SNR higher than the threshold. Out of these
simulated mergers whose signals exceed the SNR threshold (hereafter referred to as \emph{detections}),
274 for O3 and 317 for O4 have a measured value for the luminosity distance that corresponds to~\(z\leq0.2\).
We evaluate the value of V90 for each of these low-redshift events using the BAYESTAR algorithm
\citep{singer16}. For these close events, we show the cumulative distribution of V90 in Figure~\ref{fig:cdfvol}.
The blue and green histograms are detections simulated for O3 and O4, respectively. The top axis shows the
expectation value of the number of AGN within the corresponding localization volume, assuming a uniform number 
density of AGN equal to~\(n_{\rm{agn}}=10^{-4.75} \textrm{Mpc}^{-3}\). The same value for this parameter was used
in \citet{bartos17} and \citet{corley19}. This number density corresponds to AGN with a bolometric luminosity
higher than $\approx 10^{44.3}\rm{erg}\ \rm{s}^{-1}$ in the local Universe.
This value for the minimum bolometric luminosity for AGN at a specific number density has been
obtained by integrating the double power law that represents the AGN \textsc{Luminosity Function} in
\citep{hopkins07}, using the best fit values for \(z=0.1\). This holds for all the values of bolometric luminosities
mentioned hereafter.

As a sanity check, we verify that our sample of V90 from O3 simulations and the values of V90
for the \(13\) observations of O3 with redshift \(z\leq0.2\) are compatible with a single common distribution.
We do this with a 2 samples Kolmogorov-Smirnov test. We find that the hypothesis according to which the two samples
come from the same distribution cannot be rejected (p-value~\(\approx 0.39\)).


\subsection{Minimum number of GW detections to test the AGN origin}
\label{sec:estimation}

We consider a Universe where a fraction of GW events \(f_{\rm{agn}}\) originate in an AGN-type
of galaxy. Our goal is to calculate how many GW detections are needed to infer this AGN-BBH mergers connection;
more precisely, the minimum number \(N_{\rm{gw}}^{3\sigma}\) of GW detections below \(z=0.2\) needed to reject
with a \(3\sigma\) significance the hypothesis of no-connection between detected BBH mergers and AGN. We evaluate
\(N_{\rm{gw}}^{3\sigma}\) as a function of the fraction \(f_{\rm{agn}}\) of GW events originated from an AGN.
We calculate such a number by investigating the spatial correlation between AGN
positions (assumed to be uniformly distributed in comoving volume) and the localization volumes of simulated
GW detections, starting from the statistical approach presented in \citet{bartos17}.

We assume that GW localization volumes are spherical, and calculate the radius \(r_{\rm{gw}}^{\rm{max}}\)
of the biggest volume depicted in Figure \ref{fig:cdfvol}.
We then populate with AGN a sphere of radius
\begin{equation}
    r=d_\textrm{L}(0.2)+r_{\rm{gw}}^{\rm{max}}\ ,
    \label{eq:radius}
\end{equation}
where \(d_\textrm{L}(0.2)\) is the luminosity distance corresponding to \(z=0.2\). The centre of this
sphere corresponds to the position of the interferometric network we simulate the detections of.
All the AGN are treated as point sources and their distribution is uniform in comoving volume. 
We then consider a set of \(N_{\rm gw}\) GW detections and draw for each of them a value of V90
from the relevant distribution in Figure \ref{fig:cdfvol}.
We denote with \(V_i\) the localization volume associated to the \(i-th\) detection.
We require that the centre of each \(V_i\) has a distance from the interferometric network smaller
than \(d_\textrm{L}(0.2)\).
A fraction \(f_{\rm{agn}}^{\rm{eff}}=0.9f_{\rm{agn}}\) of the centres of the localization volumes
are set in order to correspond to the position of an AGN.
We here use \(f_{\rm{agn}}^{\rm{eff}}\) instead of \(f_{\rm{agn}}\) to take into account the
fact that we are here dealing with 90\% CL localization volumes, and therefore we expect only the
90\% of the origins of the simulated GWs to be actually located in such volumes. 

We then count the number \(N_i\) of AGN in each localization volume \(V_i\). Equation \ref{eq:radius}
ensures that each GW localization volume is entirely contained in our simulated Universe.

For every set of \(N_{\rm{gw}}\) simulated GW detections, we then calculate
\begin{equation}
\label{deflambda}
    \lambda=2\log\left[\frac{\mathcal{L}(f_{\rm{agn}})}{\mathcal{L}(0)}\right]\ ,
\end{equation}
where \(\mathcal{L}(0)\) and \(\mathcal{L}(f_{\rm{agn}})\) are the likelihood functions
of the no-connection hypothesis and of the \(f_{\rm{agn}}\)-correlation hypothesis, respectively.
These likelihood functions are constructed assuming a Poissonian distribution for \(N_i\).
See \citet{bartos17} for more details.

Every simulation is therefore associated to a value of \(\lambda\) that depends
on \(n_{\rm{agn}}\), \(N_{\rm{gw}}\), \(f_{\rm{agn}}\), the value of V90 of each simulated
GW event, and the number \(N_i\) of AGN within such volume.

We expect \(\lambda\) to be positive in simulations in which \(N_{\rm{gw}}f_{\rm{agn}}^{\rm{eff}}\)
localization volumes' centres correspond to an AGN. We refer to simulations that satisfy this requirement
as signal realizations, and to the value of \(\lambda\) obtained from each of them as \(\lambda_{\rm{s}}\).
Likewise, we call \(\lambda_{\rm{b}}\) every value of \(\lambda\) that is
obtained from a background realization. These realizations are simulations in which
the centres of the localization volumes are randomly distributed, uniformly in comoving volume. We, therefore,
expect \(\lambda_{\rm{b}}\) to be negative.

We perform 3,000 signal realizations and the same amount of background realizations, for each set
of values of \(N_{\rm{gw}}\), \(n_{\rm{agn}}\) and \(f_{\rm{agn}}\).

Once a value for \(f_{\rm{agn}}\) and for \(n_{\rm{agn}}\) has been set, an increase
in \(N_{\rm{gw}}\) leads to a greater separation between the distribution of
\(\lambda_{\rm{s}}\) and the distribution of \(\lambda_{\rm{b}}\).

The target degree of significance in the rejection of the no-connection hypothesis is reached
when the median value of the distribution of \(\lambda_{\rm{s}}\) corresponds to a p-value
lower than \(0.00135\) when compared to the \(\lambda_{\rm{b}}\) distribution.

To evaluate \(N_{\rm{gw}}^{3\sigma}\) for a specific value of \(f_{\rm{agn}}\),
we calculate 30 p-values, keeping such parameter fixed together with \(N_{\rm{gw}}\).
We repeat these calculations for multiple values of \(N_{\rm{gw}}\), and then fit the
trend of the average p-value for a given \(N_{\rm{gw}}\) as a function of \(N_{\rm{gw}}\)
itself. Such trend is well fitted by a decreasing exponential function for every
value of \(f_{\rm{agn}}\) we investigated.
Once the parameters of these fits are known, we invert the fit function and calculate the number
of detections corresponding to a p-value of \(0.00135\).
We repeat the same analysis for \(6\) different values of \(f_{\rm{agn}}\) between \(0.5\) and \(1\).

\vspace{-0.7em}
\section{Results}
\label{sec:res}

\subsection{Minimum number of GW detections with fixed \(n_{\rm{agn}}\)}
\label{sec:res1}

\begin{figure}
    \centering
    \includegraphics[width=0.485\textwidth]{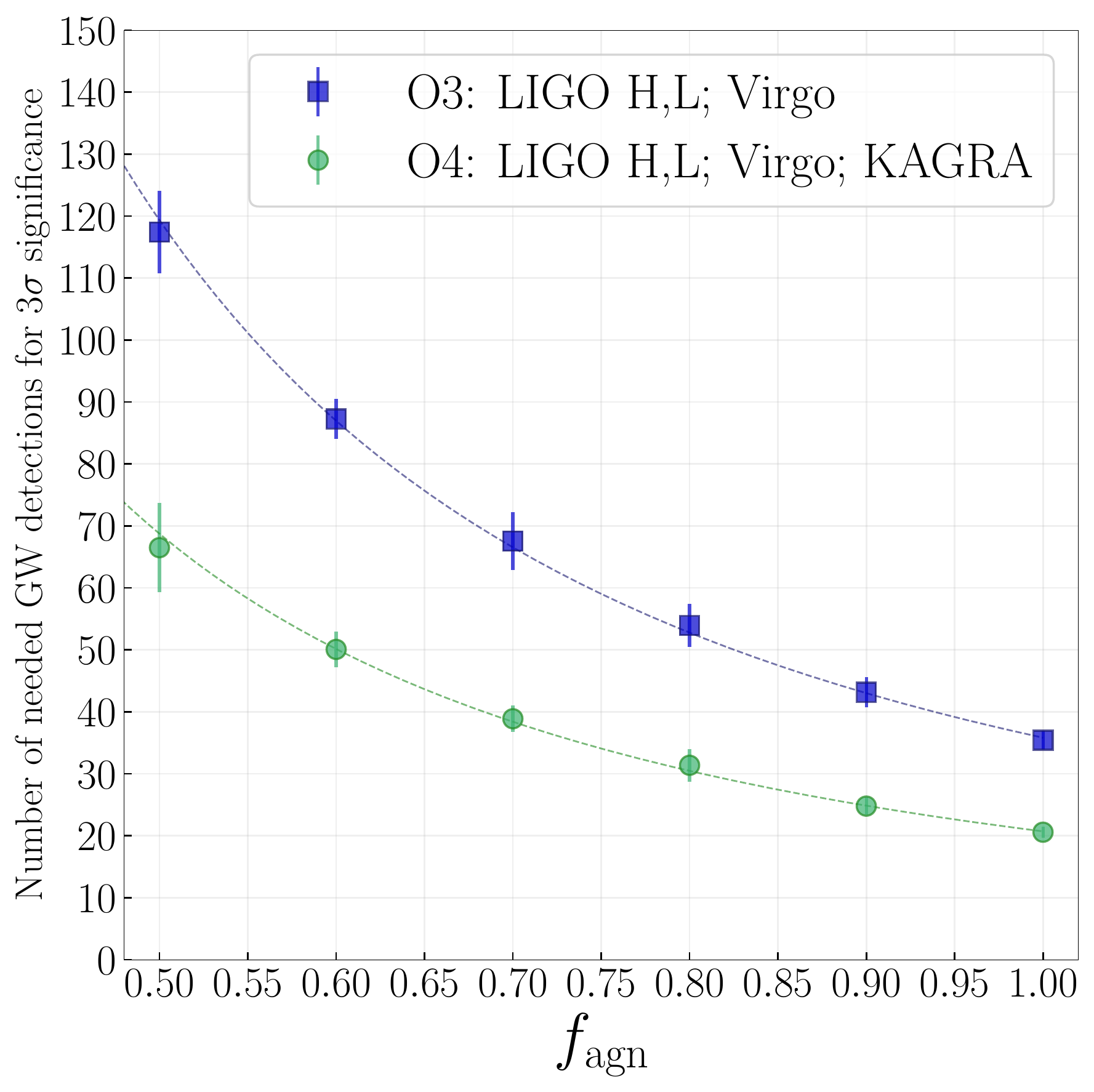}
    \caption{Number of GW detections at \(z\leq0.2\) needed to reject with a \(3\sigma\) significance
    the no GW-AGN connection hypothesis as a function of the fraction of GW originated
    from an AGN. The error bars represent the standard deviation over \(1,000\)
    realizations of \(N_{\rm{gw}}^{3\sigma}\) obtained for each tested value of \(f_{\rm{agn}}\).
    The results for the third and the fourth observing run of LVK Collaboration interferometers
    are represented by the blue squares and the green dots, respectively. The data points have
    been fitted with the following function: \(N_{\rm{agn}}^{3\sigma}=af_{\rm{agn}}^{-b}\).
    The best-fit values for the O3 scenario are \(a=35.8\pm1.2\) and \(b=1.73\pm0.08\),
    while for the O4 scenario they are \(a=20.7\pm0.7\) and \(b=1.73\pm0.11\).
    The best-fit function for O3 (O4) is represented by the blue (green) dashed line.}
    \label{fig:thresh}
\end{figure}

In this section, we present the results obtained keeping the AGN number density parameter
fixed to \(n_{\rm{agn}}=10^{-4.75}\textrm{\rm{Mpc}}^{-3}\).
The trend of \(N_{\rm{gw}}^{3\sigma}\) as a function of \(f_{\rm{agn}}\)
is shown in Figure \ref{fig:thresh}. The error bars correspond to the standard deviation of \(1,000\)
values of \(N_{\rm gw}^{3\sigma}\) calculated for each of the \(6\) values of \(f_{\rm{agn}}\) we test.
The results for O3 and O4 are represented by the blue squares and the green dots, respectively.

The trend of \(N_{\rm{gw}}^{3\sigma}\) as a function of \(f_{\rm{agn}}\) is fitted
with the same functional form used in \citet{bartos17}, which is the following:
\begin{equation}
    N_{\rm{gw}}^{3\sigma}=af_{\rm{agn}}^{-b}\ .
\end{equation}
The best fit values we obtain in the case of the O3 simulated events are 
\(a=35.8\pm1.2\) and \(b=1.73\pm0.08\), while for O4 simulated events we
obtain \(a=20.7\pm0.7\) and \(b=1.73\pm0.11\).

We perform the same analysis for O3 with a lower (\(2\sigma\)) significance
threshold for the rejection of the no-connection hypothesis. In this case, the
best-fit values for the fit are \(a=17.7\pm0.5\) and \(b=1.57\pm0.12\).
\subsection{Significance of the no-connection hypothesis rejection as a function of 
\(n_{\rm{agn}}\) and \(f_{\rm{agn}}\)}
\label{res2}

During the third observing run of the LVK Collaboration, \(13\) detected BBH mergers
have an expectation value of redshift lower than \(0.2\).
As we can infer from the results presented so far, with this low number of "closeby"
events it is not possible to reject with a \(2\sigma\) significance the
no-connection hypothesis for any value of \(f_{\rm{agn}}\), assuming
$n_{\rm{agn}}=10^{-4.75}\rm{Mpc}^{-3}$.

Nonetheless, decreasing the value of \(n_{\rm{agn}}\), every GW detection becomes more
significant, and a lower number of detection is needed to rule out the no-connection hypothesis. 

Hence, we perform the same analysis as above but keeping \(N_{\rm{gw}}\) fixed at the
value of \(13\), and varying both \(f_{\rm{agn}}\) and \(n_{\rm{agn}}\).
For each point in this 2D parameter space, we determine the p-value associated to the
median of the distribution of \(\lambda_s\) when compared to the distribution
of \(\lambda_b\).

\begin{figure*}
    \centering
    \includegraphics[width=.9\textwidth]{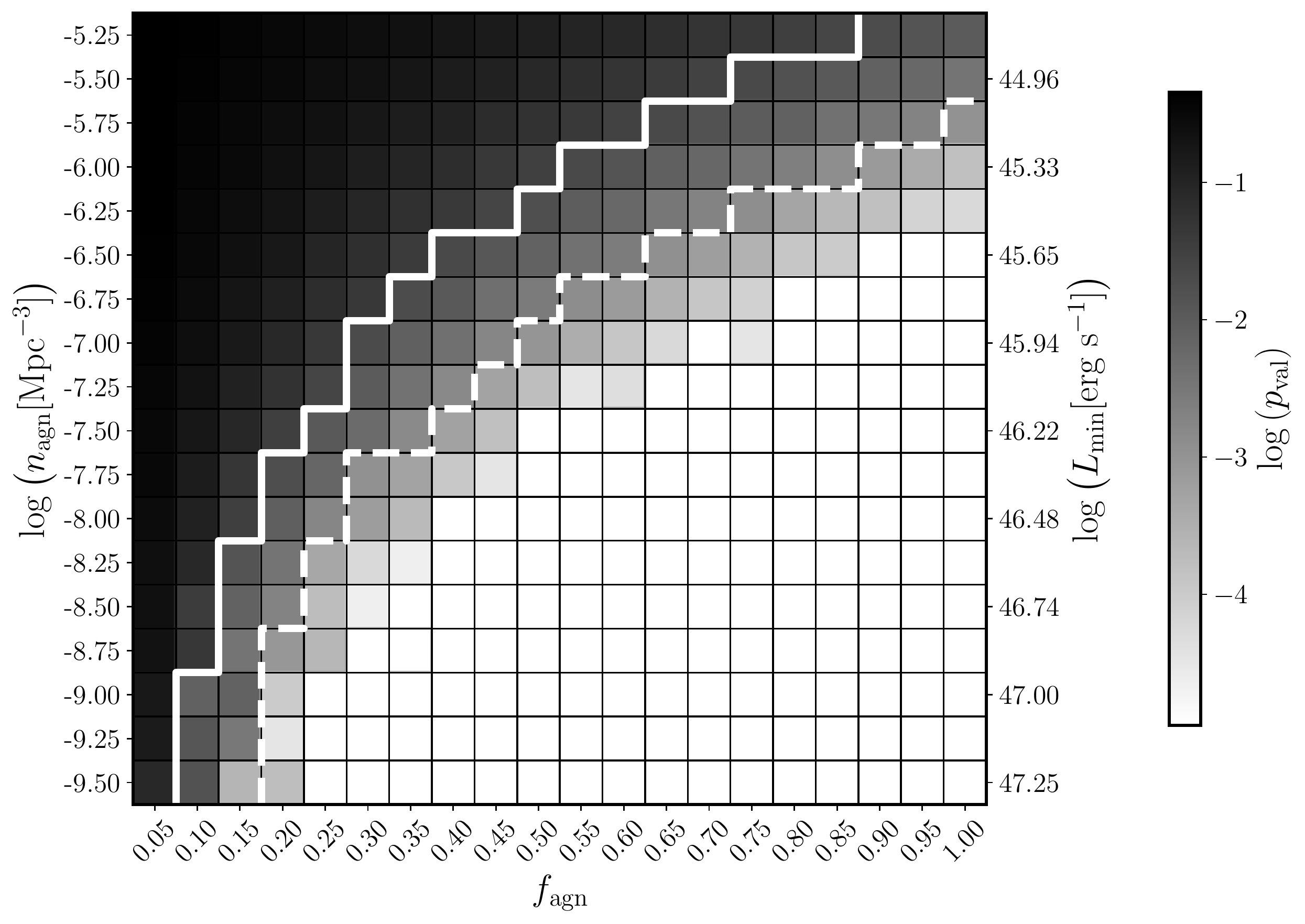}
    \caption{Significance of the rejection of the no-connection hypothesis as a 
    function of the AGN number density (\(n_{\rm{agn}}\)) and the fraction of GW
    events originated in an AGN (\(f_{\rm{agn}}\)). The p-values (and hence the significance)
    here represented refer to the detections of \(13\) events associated with \(z\leq0.2\).
    On the right of the dashed (solid) white line there is the region of the 2D parameter space in
    which the no-connection hypothesis can be rejected with a \(3\sigma\) (\(2\sigma\)) significance.
    The p-values here represented are obtained from the comparison of the median of the
    \(\lambda_{\rm{s}}\) distribution with respect to the \(\lambda_{\rm{b}}\) distribution.
    Every value of \(\lambda_{\rm{s}}\) has been calculated using Eq. \eqref{deflambda} in a simulation
    in which a fraction \(f_{\rm{agn}}\) of GWs come from an AGN. On the other hand, every value of
    \(\lambda_{\rm{b}}\) comes from a simulation in which no GW event is originated in an AGN.
    On the right-hand axis, we report the logarithm of the minimum bolometric luminosity
    \(L_{\rm{min}}[\rm{erg}\ \rm{s}^{-1}]\) that has to be considered in the integration of the AGN
    \textsc{Luminosity Function} at  \(z=0.1\) \citep{hopkins07} to obtain the value of
    \(\log(n_{\rm{agn}}[\rm{Mpc}^{-3}])\) indicated on the left-hand side of the grid.}
    \label{fig:ptab}
\end{figure*}

The results of such analysis are shown in Figure \ref{fig:ptab}. 
The white dashed (solid) line divide the parameter space into two decision
regions, corresponding to parameter choices for \(f_{\rm{agn}}\) and \(n_{\rm{agn}}\) whose
associated p-values are lower or higher than \(0.00135\) (\(0.02275\)), i.e. a significance
higher or lower than \(3\sigma\) (\(2\sigma\)), respectively. The no-connection hypothesis can
be rejected accordingly in the respective regions.

For example, with 13 GW detections and assuming a number density of AGN
\(n_{\rm{agn}}=10^{-7.50}\rm{Mpc}^{-3}\), we can, in principle, reject
the no-connection hypothesis with a \(3\sigma\) (\(2\sigma\)) significance if \(f_{\rm{agn}}\geq0.40\)
(\(f_{\rm{agn}}\geq0.25\)). Such a low number density corresponds, in the local Universe, to AGN with
bolometric luminosities \(\gtrapprox 10^{46.2}\rm{erg}\ \rm{s}^{-1}\)\citep{hopkins07}, or with central
SMBHs with masses \(\gtrapprox10^{8.5}\rm{M}_\odot\) \citep{GreeneHo07}.

\vspace{-0.5em}
\section{Discussion and conclusion}
\label{sec:concl}

\noindent We perform a statistical investigation based on the method presented in 
\citet{bartos17} in order to assess, using only AGN positions and GW localization volumes,
how many GW detections are needed to reject the no GW-AGN connection hypothesis.
We find that the \(13\) O3 GW detections with expected \(z\leq0.2\) are not enough to reject
the no-connection hypothesis with either \(3\sigma\) or \(2\sigma\) significance.
This result is obtained considering AGN with a number density \(n_{\rm{agn}}=10^{-4.75}\textrm{\rm{Mpc}}^{-3}\).
Nonetheless, Figure \ref{fig:ptab} shows that with the same number of detections, it is possible
to reject the no-connection hypothesis for specific values of the AGN number density and of
the fraction of GW events that originated inside an AGN. More precisely, the lower the AGN number
density (i.e. the higher the luminosity of the considered AGN, or the higher the mass of the central SMBH), the more likely it is to reject such
a hypothesis with a given significance threshold.
As far as O4 is concerned, the green line in Figure \ref{fig:thresh} shows that at least \(21\) detections associated
with redshift \(z\leq0.2\) will be needed to be able to reject the no-connection hypothesis between BBH mergers
and AGN with \(n_{\rm{agn}}\geq10^{-4.75}\textrm{\rm{Mpc}}^{-3}\).
The number of expected detections of BBH mergers during O4 is \(79_{-44}^{+89}\) \citep{abbott20}. In our simulations of O4
detections, roughly \(18.25\%\) of the detected events correspond to \(z\leq0.2\). Our estimate is therefore that during
O4, \(14_{-8}^{+17}\) BBH mergers will be associated with \(z\leq0.2\).
As shown in Figure \ref{fig:thresh}, \(30\) O4 closeby BBH detections would be enough to test values of \(f_{\rm{agn}}\)
higher than \(\approx 80\%\), using AGN of any luminosity. The same degree of GW-AGN connection could be tested using a lower
number of O4 detections in combination with the 13 closeby O3 detections.

We restrict our analysis to GW events with an expectation value for the redshift
of \(z\leq0.2\) for two different reasons. First, far GW events are typically associated with much larger localization
volumes than the ones associated with closer events. The inclusion of large GW localization volumes in our
algorithm makes it too computationally demanding. The second reason is that for very luminous AGN in the local Universe,
we expect to have high values of completeness in real AGN catalogues. These high values are needed in order to produce
reliable results when applying the method described in this work to real, observed GW events and AGN. The incompleteness
in observed AGN catalogues can be nonetheless taken into account with an appropriate rescaling of \(f_{\rm{agn}}\)
\citep{bartos17}.

The main assumptions we made in this work were: considering spherical GW localization volumes, and neglecting redshift
evolution for AGN and GW events as well as AGN clustering. We expect these assumptions not to remarkably impact on
our final results. The BBH merger rate, the AGN number density, and the expected AGN-assisted merger rate do not significantly
vary within the redshift range we consider \citep{hopkins07,yang20,ligo21pop}. Taking into consideration the real shape of GW events
localization volumes and the clustering of AGN in the local Universe is important when performing a maximum likelihood estimation
to find which value of \(f_{\rm{agn}}\) best represents real observations. Such estimation is not the aim of this work but is
currently being implemented in ongoing projects, in which the exploitation of realistic AGN catalogues and GW skymaps is required.

Our results motivate more in-depth statistical investigations of a possible  connection between GW events and AGN exploiting actual data. An quantitative assessment of \(f_{\rm{agn}}\) would allow us to put constrain on the BBH merger rate per AGN in 
the local Universe, that in turn can be used to both deepen our physical understanding of AGN disks and inform theoretical models of BBH mergers from the AGN formation channel. Finally, extrapolating these findings at higher redshift will enable predictions for GW events detectable by the third-generation GW interferometers.

\vspace{-0.8em}
\section*{Acknowledgements}
EMR acknowledges support from ERC Grant ``VEGA P.", number 101002511.
RB is supported by the Italian Space Agency Grant ``Phase A LISA mission activities'', Agreement No.~2017-29-H.0, CUP~F62F17000290005.
This material is based upon work supported by NSF's LIGO Laboratory which is a major facility fully funded by the
National Science Foundation. This research has made use of data or software obtained from the Gravitational Wave Open Science Center (gw-openscience.org),
a service of LIGO Laboratory, the LIGO Scientific Collaboration, the Virgo Collaboration, and KAGRA. LIGO Laboratory and
Advanced LIGO are funded by the United States National Science Foundation (NSF) as well as the Science and Technology Facilities
Council (STFC) of the United Kingdom, the Max-Planck-Society (MPS), and the State of Niedersachsen/Germany for support of
the construction of Advanced LIGO and construction and operation of the GEO600 detector. Additional support for Advanced
LIGO was provided by the Australian Research Council. Virgo is funded, through the European Gravitational Observatory (EGO),
by the French Centre National de Recherche Scientifique (CNRS), the Italian Istituto Nazionale di Fisica Nucleare (INFN)
and the Dutch Nikhef, with contributions by institutions from Belgium, Germany, Greece, Hungary, Ireland, Japan, Monaco,
Poland, Portugal, Spain. The construction and operation of KAGRA are funded by Ministry of Education, Culture, Sports, Science
and Technology (MEXT), and Japan Society for the Promotion of Science (JSPS), National Research Foundation (NRF) and Ministry
of Science and ICT (MSIT) in Korea, Academia Sinica (AS) and the Ministry of Science and Technology (MoST) in Taiwan.

{\em Software}: 
\texttt{Numpy} \citep{harris20}; 
\texttt{Matplotlib} \citep{hunter07}; 
\texttt{SciPy} \citep{virtanen20}; 
\texttt{K3Match} \citep{schellart13};
\texttt{Astropy} \citep{astropy13,astropy18};
\texttt{BAYESTAR} \citep{singer16}.

\vspace{-0.8em}
\section*{Data Availabilty}
The data underlying this article will be shared on reasonable request to the corresponding author.

\vspace{-1.0em}
\bibliographystyle{mnras}
\bibliography{bibliography}


\bsp
\label{lastpage}
\end{document}